\documentclass[12pt]{iopart}

\usepackage{cite}
\usepackage[dvips]{graphicx}
\begin{document}

\title[Penning trap robustness against entanglement]
{Robustness of spatial Penning trap modes against environment-assisted entanglement}

\author{M Genkin and A Eisfeld}

\address{Max-Planck-Institute for the Physics of Complex Systems, 01187 Dresden, Germany} 
\ead{genkin@pks.mpg.de}

\begin{abstract}
The separability of the spatial modes of a charged particle in a Penning trap in the presence of an environment is studied by means of the positive partial transpose (PPT) criterion. 
Assuming a weak Markovian environment,
described by linear Lindblad operators, our results strongly suggest that the environmental coupling of the axial and cyclotron degrees of freedom does not lead to entanglement at experimentally realistic temperatures. 
We therefore argue that,
apart from unavoidable decoherence, the presence of such an environment does not alter the effectiveness of recently suggested quantum information protocols in Penning traps, which are based on the 
combination of a spatial mode with the spin of the particle.

\end{abstract}
\submitto{\JPB}


\section{Introduction}

The possibility to confine a single charged particle in a Penning trap~\cite{Brow_86} was initially widely exploited for high precision measurements of ion masses~\cite{Blau_06,Blau_09} and the determination of 
fundamental constants~\cite{Dyck_87,Gabr_90,Blau_09_2}. More recently, 
however, Penning traps also began to attract the attention of the quantum information community, as a potential candidate for quantum computer building blocks~\cite{Marz_09,Lama_10,Gold_10}.
Indeed, the external control and cooling possibilities combined 
with the achievable high precision certainly present considerable advantages. It was shown earlier that quantum information protocols in a Penning trap are, at least theoretically, possible; The schemes 
known so far are based on information storage in either the axial or the cyclotron degree of freedom and the spin of the particle~\cite{Manc_99,Ciar_03,Ciar_04}. Hence, as in any potential qubit 
implementation, one has to address the effects of an environment on the device.
While decoherence is to some extent unavoidable, in the particular case of a Penning trap the environment may have yet another undesirable effect, as it allows for coupling of degrees of freedom which 
would not occur otherwise. More specific, the axial and the cyclotron motion are no longer uncoupled if an environment is present, and consequently the separability of the latter cannot be taken for granted.
It is the purpose of the present work to examine whether environment-assisted entanglement of these modes is likely to emerge. 


The intended analysis essentially requires a model which incorporates environmental effects into the dynamics and a measure for separability (or entanglement) which can be extracted from such a model. 
For simplicity, we will assume a Markovian environment, and treat it by means of a standard master equation of Lindblad type~\cite{Lind_76}. At this point it should be mentioned that generally both a Markovian 
and a non-Markovian bath can (but not necessarily do) lead to environment-assisted entanglement~\cite{Isar_09,Hoer_08,Li_10,Paz_08,Corm_10}, so that this particular choice should not rule out an outcome {\it a priori}. 
Assuming linear coupling to the environment, the equations of motion for a charged particle in a Penning remain analytically solvable. Furthermore, since the Penning trap Hamiltonian is 
at most quadratic in the canonical variables, initially Gaussian states remain Gaussian for all times. For Penning traps, such class of Gaussian
coherent states was recently derived~\cite{Fern_09,Asto_11}. This makes it possible to establish a continuous variable separability criterion for the modes which is completely determined by the covariance matrix,
as the general PPT criterion greatly simplifies for a system of 1 vs. $N-1$ symmetric Gaussian modes. We will first describe the environment model in Section~\ref{sec2}, from which the time evolution of the covariance matrix
is obtained. We will then introduce a separability criterion which is extractable from the covariance matrix in Section~\ref{secsep}.
The results and conclusions will be presented in Section~\ref{seccalc}. The work is summarized in Section~\ref{secsum}.

\section{Phenomenological modelling of the environment}\label{sec2}
The characteristic frequencies of a charged particle with mass $m$ and charge $q$ in a Penning trap are determined by the trap parameters (magnetic field $B$, trap dimension $d$ and electrode voltage $U$)
as follows:
\begin{equation}
\omega_c= \frac{qB}{m}\,\,({\rm cycltotron}),\quad \omega_z=\sqrt{\frac{|q|U}{md^2}} \,\,({\rm axial}),
\end{equation}
and the radial frequency is defined as
\begin{equation}
\omega_{\perp}^2=\frac{\omega_c^2}{4}-\frac{\omega_z^2}{2}.
\end{equation}
The corresponding Hamiltonian can be written as
\begin{equation}
H=H_{\perp}+H_z,
\end{equation}
with
\begin{eqnarray}
H_{\perp}=\frac{p_x^2+p_y^2}{2m}+\frac{\omega_c}{2}(xp_y-yp_x)+\frac{m\omega_{\perp}^2}{2}\left(x^2+y^2\right),\\
H_z=\frac{p_z^2}{2m}+\frac{1}{2}m\omega_z^2z^2,
\end{eqnarray}
where the additional spin term is omitted, since the spin motion is completely separable
from the dynamics and does not affect the calculations. We introduce the phase space vector 
\begin{equation}
{\bf R}=(x,p_x,y,p_y,z,p_z)^{\rm T}
\end{equation}
and the covariance matrix of the system
\begin{equation}\label{CM}
\sigma_{ij}=\frac{1}{2}{\rm Tr}\left[\rho(R_iR_j+R_jR_i)\right]-{\rm Tr}(\rho R_i){\rm Tr}(\rho R_j),\quad i,j=1,\dots,6
\end{equation}
where $\rho$ is the reduced density operator of the system. Its time evolution in the presence of a weak Markovian environment can be described by a master equation of the form
\begin{equation}\label{master}
\frac{\rmd\rho}{\rmd t}=-\frac{i}{\hbar}[H,\rho]+\frac{1}{2\hbar}\sum_{j}\left(\left[V_{j}\rho,V_{j}^{\dagger}\right]+\left[V_{j},\rho
V_{j}^{\dagger}\right]\right)
\end{equation}
with Lindblad operators $V_j$ linear in the operators $p_k$ and $x_k$
\begin{equation}
V_j=\sum_{k=1}^3(a_{jk}p_k+b_{jk}x_k),\quad V_j^{\dagger}=\sum_{k=1}^3(a_{jk}^*p_k+b_{jk}^*x_k),\quad  j=1,..,6
\end{equation}
where $a_{jk},b_{jk}$ are complex coefficients.
Models of this type are well known~\cite{Sand2_87,Sand_87,Isar_94}, and such an approach was adopted earlier to study damping and decoherence behavior in Penning traps~\cite{Genk_09}. 
However, unlike therein, here we will not simplify the picture by demanding that the damping in a certain degree of freedom may not affect the
dynamics of another degree of freedom.
By transforming the master equation to the Heisenberg picture, one obtains the following time evolution of the covariance matrix:
\begin{equation}\label{sol2}
\sigma(t)=\exp(\Lambda t)(\sigma(0)-\Gamma)(\exp(\Lambda t))^{\rm T}+\Gamma,
\end{equation}
where $\sigma(0)$ is the covariance matrix of the initially prepared state. The $6\times 6$ matrices $\Lambda$ and $\Gamma$ are
completely determined by the Lindblad coefficients $a_{jk},b_{jk}$ and the parameters of the Hamilton operator. 
As suggested in~\cite{Sand_87}, it is convenient to introduce the vectors
\begin{equation}
{\bf a}_k=(a_{1k},..,a_{6k})^{\rm T},\quad {\bf b}_k=(b_{1k},..,b_{6k})^{\rm T},\quad k=1,2,3,
\end{equation}
with the scalar product
\begin{equation}
\langle {\bf f},{\bf g}\rangle=\sum_{i=1}^6 f_i^*g_i,
\end{equation}
and the phenomenological constants ($k,l=1,2,3$)
\numparts
\begin{eqnarray}\label{phen1}
\alpha_{kl}=-\alpha_{lk}=-{\rm Im}\left(\langle {\bf a_k},{\bf a_l}\rangle\right),\\
\beta_{kl}=-\beta_{lk}=-{\rm Im}\left(\langle {\bf b_k},{\bf b_l}\rangle\right),\\
\lambda_{kl}=-{\rm Im}\left(\langle {\bf a_k},{\bf b_l}\rangle\right),\\
\label{D1}
D_{x_kx_l}=D_{x_lx_k}=\frac{\hbar}{2}{\rm Re}\left(\langle {\bf a_k},{\bf a_l}\rangle\right),\\
\label{D2}
D_{p_kp_l}=D_{p_lp_k}=\frac{\hbar}{2}{\rm Re}\left(\langle {\bf b_k},{\bf b_l}\rangle\right),\\
\label{phenlast}
D_{x_kp_l}=D_{p_lx_k}=-\frac{\hbar}{2}{\rm Re}\left(\langle {\bf a_k},{\bf b_l}\rangle\right).
\end{eqnarray}
\endnumparts
The matrix $\Lambda$ is then given by
\begin{equation}\label{Lambda}\fl
\Lambda=\left(\begin{array}{cccccc}
-\lambda_{11} & 1/m & -\lambda_{12}-\omega_c/2 & -\alpha_{12} & -\lambda_{13} & -\alpha_{13} \\
-m\omega_{\perp}^2 & -\lambda_{11} & \beta_{12} & -\lambda_{21}-\omega_c/2 & \beta_{13} & -\lambda_{31} \\
-\lambda_{21}+\omega_c/2 & \alpha_{12} & -\lambda_{22} & 1/m & -\lambda_{23} & -\alpha_{23} \\
-\beta_{12} & -\lambda_{12}+\omega_c/2 & -m\omega_{\perp}^2 & -\lambda_{22} & \beta_{23} & -\lambda_{32} \\
-\lambda_{31} & \alpha_{13} & -\lambda_{32} & \alpha_{23} & -\lambda_{33} & 1/m \\
-\beta_{13} & -\lambda_{13} & -\beta_{23} & -\lambda_{23} & -m\omega_{z}^2 & -\lambda_{33}
\end{array}\right)
\end{equation}
and the symmetrical matrix $\Gamma$ is the solution of the linear equation system~\cite{Sand_87} 
\begin{equation}\label{lineq}
\Lambda\Gamma+\Gamma\Lambda^{T}=-2D.
\end{equation}
Here, $D$ denotes the symmetric $6\times 6$ diffusion matrix, which is defined as $D_{ij}=D_{R_iR_j}$. 
It is also important to note that the phenomenological constants defined above cannot be chosen arbitrarily but have to satisfy
certain conditions in order to preserve the positivity of the density matrix and the uncertainty relation. These restrictions are summarized in the appendix.

\section{Separability criterion for Gaussian states}\label{secsep}
A necessary condition for the separability of a bipartite system was formulated in the 1990s by Peres~\cite{Pere_96} and Horodecki {\it et al}~\cite{Horo_96}.
They have proven that this information is directly related to the positivity of the partial transpose (PPT) of the density matrix with respect to one of the
partitions. Although this is a powerful criterion, it is not always easliy applicable in practice in its original form. However, Gaussian multimode states
have the unique feature that the complete information about the state can be extracted from the corresponding first two moments (i.e. the expectation values and the
covariance matrix), without the necessity to deal explicitly with the density matrix. For practical purposes, this property is very valuable;
In particular, it can be shown that the full uncertainty relation as well as conditions for separability or entanglement are extractable from the covariance matrix alone.
Recent reviews with focus on entanglement in continuous variable systems can be found e.g. in~\cite{Brau_05,Ades_07}. Below, we only
briefly summarize the most crucial aspects relevant for the present study.

\begin{enumerate}
\item{\it Uncertainty relation}\newline
Let us consider a system of $N$ Gaussian modes, with the phase space vector ${\bf R}^N=(x_1,p_1,..,x_N,p_N)$. We define the symplectic matrix $\Omega$ as
\begin{equation}\label{Omega}
\Omega_{jk}=\frac{1}{i}\left[R^N_j,R^N_k\right].
\end{equation}
Then the covariance matrix $\sigma$, defined as in~(\ref{CM}), has to satisfy the following so called {\it bona fide} condition in order to preserve the full uncertainty relation~\cite{Simo_94}:
\begin{equation}
\sigma+\frac{i}{2}\Omega\geq 0.
\end{equation}
In other words, only covariance matrices satisfying the above condition are physically meaningful. For the particular case of a Penning trap, some related aspects were discussed in~\cite{Hacy_96}.
\item{\it Separability}\newline
Given a covariance matrix $\sigma$ that satisfies the {\it bona fide} condition, the following necessary and sufficient condition
for separability holds for a system of $m+n$ bisymmetric modes~\cite{Simo_00,Wern_01,Sera_06}: 
\begin{equation}\label{entcond}
\tilde{\sigma}+\frac{i}{2}\Omega\geq 0.
\end{equation}
Here, $\tilde{\sigma}$ is the covariance matrix of the partially transposed state with respect to either of the partitions. The partial transposition
in phase space corresponds to an operation which switches the sign of the momenta belonging to a partition, hence $\tilde{\sigma}$ is given by
\begin{equation}\label{ptrans}
\tilde{\sigma}=W\sigma W,\quad W=\oplus_{1}^ms_z\oplus 1_{2n},
\end{equation}
where $s_z$ denotes the Pauli $z$-matrix and $1_{2n}$ is the $2n\times 2n$ unity matrix.
We emphasize that this holds, in particular, for the special case $m=N-1$, $n=1$ which will be considered below.
\end{enumerate}
Given the above criteria, the time evolution of the covariance matrix~(\ref{sol2}) thus provides us with the information whether environment-assisted entanglement 
occurs at any given time. This is the case whenever the condition~(\ref{entcond}) is violated. 

\section{Calculations, results and conclusions}\label{seccalc}
\subsection{Preliminary remarks}
Following the notation from Eqs.~(\ref{Omega})-(\ref{ptrans}), the Penning trap is a particular case with $N=3$, $m=2$ and $n=1$, since the cyclotron motion
is completely symmetric in the $x$ and $y$ modes. The partial transposition  with respect to the axial mode therefore 
becomes $W={\rm diag}(1,-1,1,-1,1,1)$ and is applied to the covariance matrix defined in~(\ref{CM}). Since the matrix $\tilde{\sigma}+\frac{i}{2}\Omega$ is Hermitean,
it is sufficient to compute its lowest eigenvalue $\epsilon$ in order to monitor the (violation of) non-negativity. 
The initial covariance matrix $\sigma_0$, which enters the time evolution~(\ref{sol2}), was calculated for the Penning trap coherent states in~\cite{Fern_09,Genk2_09}.
The only remaining problem is to find an appropriate choice for the phenomenological constants~(\ref{phen1})-(\ref{phenlast}), which is indeed one of the major obstacles when working with master
equations of the form~(\ref{master}). In the present case, the number of these parameters can be reduced for symmetry reasons; Furthermore, to some of the remaining constants
it is rather straightforward to assign a physical meaning. Still, the phenomenology is not fully resolved. In what follows, we discuss this problem and suggest a 
Monte-Carlo-like approach to circumvent it.
\begin{enumerate}
\item{\it Symmetry considerations and damping rates}\newline
To illustrate the meaning of the elements of the matrix $\Lambda$ defined in~(\ref{Lambda}), we note that the equations of motion for the first moments (i.e. expectation values) derived from the master equation
read
\begin{equation}
\frac{\rmd}{\rmd t}\langle {\bf R}\rangle (t)=\Lambda \langle {\bf R}\rangle (t),
\end{equation}
with the solution 
\begin{equation}
\langle {\bf R}\rangle (t)=\exp(\Lambda t)\langle {\bf R}\rangle (0).
\end{equation}
We thus see that for $k\neq l$ the parameters $\lambda_{kl}$ couple the coordinates $x_k$ and $x_l$ and the momenta $p_k$ and $p_l$, while
the parameters $\alpha_{kl}$, $\beta_{kl}$ couple the coordinate $x_k$ with the momentum $p_l$. Since the system at hand is completely symmetric in $x$ and $y$,
we can reduce the number of parameters by setting
\begin{eqnarray}\nonumber
\lambda_{12}=\lambda_{21},\quad \lambda_{13}=\lambda_{23},\quad \lambda_{31}=\lambda_{32},\\
\label{phenokonst}
\alpha_{13}=\alpha_{23},\quad \beta_{13}=\beta_{23}.
\end{eqnarray} 
Moreover, for $k=l$ the constants $\lambda_{kk}$ are identified as damping rates for the mode $k$, and, again for symmetry reasons, we can set $\lambda_{11}=\lambda_{22}$.
\item{\it Diffusion coefficients and temperature}\newline
For the same symmetry reasons as above, the number of independent diffusion coefficients~(\ref{D1})-(\ref{phenlast}) can be reduced by setting
\begin{eqnarray}\nonumber
D_{xx}=D_{yy},\quad D_{xz}=D_{yz},\quad D_{p_xp_x}=D_{p_yp_y},\\
D_{p_xp_z}=D_{p_yp_z},\quad D_{xp_y}=D_{yp_x},\quad D_{xp_x}=D_{yp_y},\\
\nonumber
D_{xp_z}=D_{yp_z},\quad D_{zp_x}=D_{zp_y}.
\end{eqnarray}
In addition, the temperature $T$ of the environment can be incorporated by the following choice for the diagonal elements of the diffusion matrix:
\begin{eqnarray}\label{DTemp}\fl
D_{xx}=D_{yy}=\frac{\hbar\lambda_{11}}{2m\omega_{\perp}}\coth\left(\frac{\hbar\omega_{\perp}}{2k_BT} \right),\quad D_{zz}=\frac{\hbar\lambda_{33}}{2m\omega_{z}}\coth\left(\frac{\hbar\omega_{z}}{2k_BT} \right),\\
\nonumber
\fl
D_{p_xp_x}=D_{p_yp_y}=\frac{\hbar\lambda_{11}m\omega_{\perp}}{2}\coth\left(\frac{\hbar\omega_{\perp}}{2k_BT} \right),\quad D_{p_zp_z}=\frac{\hbar\lambda_{33}m\omega_{z}}{2}\coth\left(\frac{\hbar\omega_{z}}{2k_BT} \right),
\end{eqnarray}
$k_B$ being the Boltzmann constant.
This can be seen as a multidimensional extension of diffusion coefficients corresponding to an asymptotic Gibbs state. The choice is well-known in one-dimensional models of such type~\cite{Sand2_87,Isar_94,Isar_00,Palc_00}.

\item{\it Monte-Carlo approach}\newline
Since the exact form of the environment is not known {\it a priori}, 
the computation of the lowest eigenvalue $\epsilon$ of $\tilde{\sigma}+i\Omega/2$ as a function of time for some particular choice of the remaining independent phenomenological constants would not really answer the question whether environment-assisted
entanglement emerges; Instead, a scheme is implemented where a large number of trajectories $\epsilon(t)$ is computed with the independent environmental constants being chosen randomly for each trajectory. 
The random generation is, however, restricted by the Born-Markov condition underlying the master equation~(\ref{master}). This is taken into account when choosing the intervals from which the random variables are selected.
In addition, the constraints listed in the appendix have to be satisfied for each trajectory. If this is not the case, the trajectory is discarded. The same applies if a violation of the {\it bona fide} condition
occurs. The remaining physically meaningful trajectories $\epsilon(t)$ obtained in this way allow us an insight into the entanglement dynamics and its dependence on the bath temperature.

\end{enumerate}
 
\subsection{Exemplary calculations and discussion}
As an example, we consider a proton in a Penning trap with the following parameters:
\begin{equation}
\omega_c\approx 484\,{\rm MHz},\quad\omega_z\approx 63.2\,{\rm MHz}.
\end{equation}
The values are taken from Table II in~\cite{Brow_86}. From here on, a system of units is used in which we set $\hbar$, the proton mass and $\omega_z$ equal to unity.
The positive damping rates $\lambda_{11}=\lambda_{22}$ and $\lambda_{33}$ are generated as uncorrelated, uniformly distributed random numbers in the interval $[10^{-2},10^{-1}]$,
thus obeying the Markovian assumption $\lambda_{kk}\ll\omega_c,\,\omega_z$
while the coupling constants $\lambda_{kl},\,\alpha_{kl},\,\beta_{kl}$  with $k\neq l$ were assumed to 
be even weaker than the direct coupling terms and generated in the same manner from the manifold $[10^{-3},10^{-2}]\bigcup\,[-10^{-2},-10^{-3}]$. 
To fulfill the conditions~(\ref{append}) we have for the off-diagonal elements of the diffusion matrix:
\begin{eqnarray}\nonumber
D_{x_kx_l}=\xi_{x_kx_l}\sqrt{D_{x_kx_k}D_{x_lx_l}-\frac{\hbar^2\alpha_{kl}^2}{4}},\\
D_{p_kp_l}=\xi_{p_kp_l}\sqrt{D_{p_kp_k}D_{p_lp_l}-\frac{\hbar^2\beta_{kl}^2}{4}},\\
\nonumber
D_{x_kp_l}=\xi_{x_kp_l}\sqrt{D_{x_kx_k}D_{p_lp_l}-\frac{\hbar^2\lambda_{kl}^2}{4}},
\end{eqnarray}
where $\xi_{R_kR_l}\in [-1,1]$ are a set of random numbers.
 
\begin{figure}
\begin{footnotesize}
\begin{center}
\scalebox{0.85}{\includegraphics{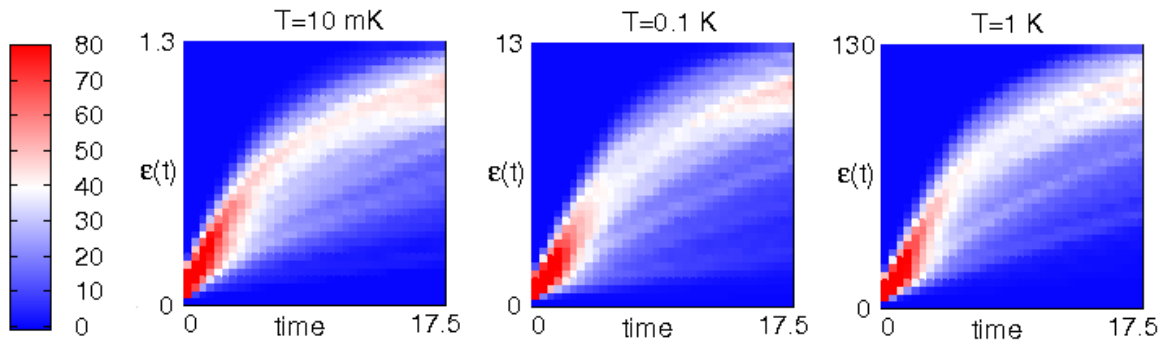}}\\
\end{center}
\begin{quote}
\caption{(Color online) Monte-Carlo trajectories of the lowest eigenvalue $\epsilon$ of the matrix $\tilde{\sigma}+i\Omega/2$, where  $\tilde{\sigma}$ is the partially transposed covariance matrix, shown as a function
of time for three different temperatures. The coloring indicates the number of trajectories passing through a given bin; We particularly emphasize the absence of negative values of $\epsilon$ (all trajectories start at $\epsilon(t=0)=0$), which persists even for
$10^{6}$ computed trajectories (not shown here). The time unit in the chosen unit system is $\approx 15.8\,$ns.}
\label{fig1}
\end{quote}
\end{footnotesize}
\end{figure}

\begin{figure}
\begin{footnotesize}
\begin{center}
\scalebox{0.99}{\includegraphics{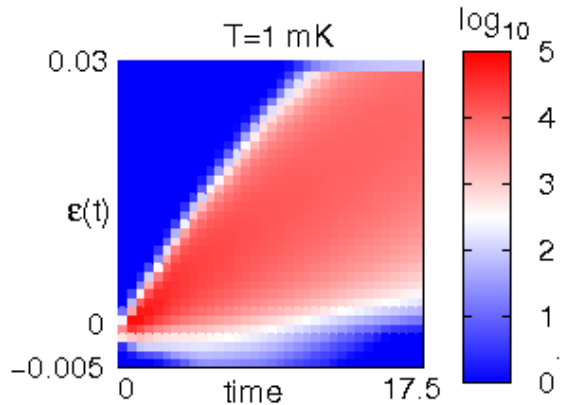}}\\
\end{center}
\begin{quote}
\caption{(Color online) Same plot as in Figure 1 but at $T=1\,$mK, shown for $10^7$ trajectories on a logarithmic scale (however, for a better visualisation the shading (color) of the bins containing zero trajectories is set to zero instead of $-\infty$).
At this low temperature, the emergence of entanglement is clearly visible.}
\label{fig2}
\end{quote}
\end{footnotesize}
\end{figure}

Figure~\ref{fig1} shows the time evolution of the lowest eigenvalue $\epsilon$ of $\tilde{\sigma}+i\Omega/2$, computed for three different temperatures (10 mK, 0.1 K and 1 K).
Each of the plots contains 1000 trajectories (less the discarded non-physical ones).
The diagrams were dissected into single bins, where the coloring indicates the number of trajectories passing through a given bin. 
We note that qualitatively the results are nearly identical, while quantitatively $\epsilon$ scales linearly with temperature. The linear scaling is a typical sign of the
high-temperature limit, as one obtains from the Taylor expansion of the diffusion coefficients~(\ref{DTemp}) $\coth\left(\hbar\omega/(2k_BT)\right)\sim T$ for $2k_BT\gg\hbar\omega$.  
For the temperatures considered, the latter condition does indeed hold, and hence one would expect the detrimental effect of thermalization on entanglement to be dominant compared to the weak environmental coupling of the modes.
The results shown in Figure~\ref{fig1} confirm the expectation - from the diagrams it is evident that, even if possible, in the studied case environment-assisted entanglement is very unlikely, as for all the trajectories there is not 
a single event of $\epsilon$ becoming negative at any time. We note by passing that the same observation holds also if the number of trajectories is increased by 3-4 orders of magnitude,
which just leads to a rescaling of the plots, leaving the number of entanglement events unchanged equal to zero. Thus, at the typical operation temperatures, the environment is much more likely to destroy rather than create entanglement in a Penning trap.
In order to corroborate this hypothesis, we also considered the very low temperature case $T=1\,$mK, which, to the best of our knowledge, is not yet experimentally feasible for a Penning trap. In this case the high-temperature limit breaks down 
and consequently the thermalization is less severe, thus possibly allowing for environment-assisted entanglement. Indeed, further numerical simulations support this conclusion. Figure~\ref{fig2} shows $10^7$ trajectories (less the discarded non-physical ones),
computed in the same manner as those displayed in Figure~\ref{fig1} but at $T=1\,$mK. One clearly observes that the non-negativity of $\tilde{\sigma}+i\Omega/2$ is violated for some trajectories, which is a signature of entanglement.
To ensure that the entanglement is solely due to the temperature decrease and not a particular constellation of random numbers, we also checked that the sets of random numbers that give rise to entanglement at low temperatures do not do so 
at higher temperatures. Therefore, we can conclude that our results strenghten the assumption of separability of the cyclotron and axial mode in quantum infromation related applications of Penning traps,
at least in the case of a weak Markovian environment at realistic temperatures.

\section{Summary}\label{secsum}
We investigated the possibility of environment-assisted entanglement of the spatial modes of a single charged particle in a Penning trap. The system-environment interaction was
modelled by a Markovian master equation, with Lindblad operators preserving the Gaussian form of an initially prepared Penning trap coherent state. This allowed us to apply a specific practical
form of the PPT separability criterion extractable from the covariance matrix, the time evolution of which is analytically solvable in the framework of the adopted model. 
The generally unknown phenomenological parameters of the master equation were treated by means of a Monte-Carlo-like simulation. We found that
the environment is very unlikely to cause entanglement in the considered case, implying robust separability of spatial Penning trap modes. By comparing the time evolution
for different temperatures, this behavior should most likely be attributed to the generally destructive effect of thermalization on entanglement. This is supported by entanglement signatures
occuring at unrealisticly low temperatures. Thermalization thus turns out to be much more significant than the weak coupling of the modes caused by environmental scattering.

{\it Note added:} After finishing the manuscript, we became aware of two very recent papers which should be mentioned in connection with the present work. The first paper by Hamdouni~\cite{Hamd_10} contains an explicit analytical derivation of multidimensional diffusion coefficients for the same
environment model as used here. This result links the diffusion coefficients to the elements of $\Lambda$ (Eq.~(\ref{Lambda})), which in principle allows us to restrict the Monte-Carlo approach solely to the phenomenological constants given in~(\ref{phenokonst}). 
The second paper by Anastopoulos {\it et al}~\cite{Anas_10} addresses the generalized uncertainty relations and generation of entanglement in quantum Browninan motion using Wigner propagator techniques; our results qualitatively agree with several findings of their work.

\section*{Acknowledgment}
We thank Clemens Gneiting for proofreading the manuscript.

\appendix
\section{Restrictions on the phenomenological constants}
The constraints are a straightforward three-dimensional extension of the relations known for the two-dimensional case~\cite{Sand_87}. The parameters have to be chosen such that
the principal minors of the following matrix are non-negative:
$$
\left(\begin{array}{cccccc}
D_{xx} & D_{xy}-\frac{i\hbar\alpha_{12}}{2} &  D_{xz}-\frac{i\hbar\alpha_{13}}{2} & -D_{xp_x}-\frac{i\hbar\lambda_{11}}{2} & -D_{xp_y}-\frac{i\hbar\lambda_{12}}{2} & -D_{xp_z}-\frac{i\hbar\lambda_{13}}{2} \\
D_{xy}+\frac{i\hbar\alpha_{12}}{2} & D_{yy} & D_{yz}-\frac{i\hbar\alpha_{23}}{2} & -D_{yp_x}-\frac{i\hbar\lambda_{21}}{2} & -D_{yp_y}-\frac{i\hbar\lambda_{22}}{2} & -D_{yp_z}-\frac{i\hbar\lambda_{23}}{2} \\
D_{xz}+\frac{i\hbar\alpha_{13}}{2} & D_{yz}+\frac{i\hbar\alpha_{23}}{2} & D_{zz } & -D_{zp_x}-\frac{i\hbar\lambda_{31}}{2} & -D_{zp_y}-\frac{i\hbar\lambda_{32}}{2} & -D_{zp_z}-\frac{i\hbar\lambda_{33}}{2} \\
-D_{xp_x}+\frac{i\hbar\lambda_{11}}{2} & -D_{yp_x}+\frac{i\hbar\lambda_{21}}{2} & -D_{zp_x}+\frac{i\hbar\lambda_{31}}{2} & D_{p_xp_x} & D_{p_xp_y}-\frac{i\hbar\beta_{12}}{2} & D_{p_xp_z}-\frac{i\hbar\beta_{13}}{2} \\
-D_{xp_y}+\frac{i\hbar\lambda_{12}}{2} & -D_{yp_y}+\frac{i\hbar\lambda_{22}}{2} & -D_{zp_y}+\frac{i\hbar\lambda_{32}}{2} & D_{p_xp_y}+\frac{i\hbar\beta_{12}}{2} & D_{p_yp_y} & D_{p_yp_z}-\frac{i\hbar\beta_{23}}{2} \\
-D_{xp_z}+\frac{i\hbar\lambda_{13}}{2} & -D_{yp_z}+\frac{i\hbar\lambda_{23}}{2} & -D_{zp_z}+\frac{i\hbar\lambda_{33}}{2} & D_{p_xp_z}+\frac{i\hbar\beta_{13}}{2}  & D_{p_yp_z}+\frac{i\hbar\beta_{23}}{2} & D_{p_zp_z}
\end{array}\right). $$
In particular, the definitions~(\ref{phen1})-(\ref{phenlast}) imply, because of the Cauchy-Schwarz inequality, the following constraints ($k,l=1,2,3$):
\begin{eqnarray}\nonumber
D_{x_kx_k}D_{x_lx_l}-D_{x_kx_l}^2\geq\frac{\hbar^2\alpha_{kl}^2}{4},\\
\label{append}
D_{p_kp_k}D_{p_lp_l}-D_{p_kp_l}^2\geq\frac{\hbar^2\beta_{kl}^2}{4},\\ 
\nonumber
D_{x_kx_k}D_{p_lp_l}-D_{x_kp_l}^2\geq\frac{\hbar^2\lambda_{kl}^2}{4}.
\end{eqnarray}
Finally, we note that the system only approaches an asymptotic state if the matrix $\Lambda$ defined in~(\ref{Lambda}) has no eigenvalues with positive real parts. In the illustrative Monte Carlo calculations,
all trajectories $\epsilon(t)$ computed with a set of parameters violating any of the above conditions were discarded.

\section*{References}
\bibliographystyle{unsrt.bst}
\bibliography{ent.bib}
\end{document}